\documentclass[graphics,floatfix, nofootinbib,tightenlines,nobibnotes, 12pt]{revtex4-1}

\usepackage{graphicx}
\usepackage[english]{babel}
\usepackage[utf8]{inputenc}
\usepackage{amsmath,amssymb}
\usepackage{amsfonts}
\usepackage{multirow}
\usepackage{pstricks}
\usepackage{float}
\usepackage{subfigure}
\usepackage{color}
\usepackage{epsfig}
\usepackage{slashed}
\usepackage[colorlinks=true, linkcolor=blue, citecolor=blue, urlcolor=blue]{hyperref}

\begin{document}

\title{Production of semi-inclusive doubly heavy baryons via top-quark decays}

\author{Juan-Juan Niu$^{1}$}
\email{niujj@cqu.edu.cn}
\author{Lei Guo$^{1}$}
\email{guoleicqu@cqu.edu.cn, corresponding author}
\author{Hong-Hao Ma$^{2}$}
\email{mahonghao.br@gmail.com}
\author{Xing-Gang Wu$^{1}$}
\email{wuxg@cqu.edu.cn}
\author{Xu-Chang Zheng$^{1}$}
\email{zhengxc@cqu.edu.cn}

\address{$^{1}$ Department of Physics, Chongqing University, Chongqing 401331, People's Republic of China}
\address{$^{2}$ Faculdade de Engenharia de Guaratinguet\'a, Universidade Estadual Paulista, 12516-410, Guaratinguet\'a, S\~{a}o Paulo, Brazil}

\date{\today}

\begin{abstract}

In the paper, we present a detailed discussion on the semi-inclusive production of doubly heavy baryons ($\Xi_{bc}$ and $\Xi_{bb}$) through top-quark decay channel, $t\rightarrow \Xi_{bQ^{\prime}}+ \bar {Q^{\prime}} + W^+ $, within the framework of nonrelativistic QCD. In our calculations, the contributions from the intermediate diquark states, $\langle bc\rangle[^{3}S_{1}]_{\mathbf{\bar 3}/ \mathbf{6}}$, $\langle bc\rangle[^{1}S_{0}]_{\mathbf{\bar 3}/ \mathbf{6}}$, $\langle bb\rangle[^{1}S_{0}]_{\mathbf{6}}$, and $\langle bb\rangle[^{3}S_{1}]_{\mathbf{\bar 3}}$, have been taken into consideration. Main uncertainties from the heavy quark mass $(m_c,~m_b~\mbox{or}~m_t)$, the renormalization scale $\mu_r$, and the nonperturbative transition probability have been estimated. For a comparison, we also analyze the production of doubly heavy baryons under the approximate fragmentation function approach. Estimated at the LHC or a High Luminosity LHC with $\mathcal{L}$ =$10^{34-36}~\rm{cm}^{-2}~\rm{s}^{-1}$, there will be about $2.25\times10^{4-6}$ events of $\Xi_{bc}$ and $9.49\times10^{2-4}$ events of $\Xi_{bb}$ produced in one operation year through top-quark decays.


\end{abstract}

\maketitle

\section{Introduction}

Recently, the LHCb Collaboration of the CERN LHC announced its first observation of the doubly heavy baryon $\Xi_{cc}^{++}$~\cite{Aaij:2017ueg} for the proton-proton center-of-mass collision energy of $13$~TeV and an integrated luminosity of 1.7~$fb^{-1}$, confirming the success of the quark model~\cite{GellMann:1964nj, Zweig:1981pd, Zweig:1964jf, DeRujula:1975qlm}. However, there is still no explicit evidence on the doubly heavy baryons $\Xi_{bc}$ and $\Xi_{bb}$ (throughout the paper, we label them as $\Xi_{bQ'}$, with $Q^{\prime}$ representing the heavy $b$ or $c$ quark, respectively). A careful study on the $\Xi_{bQ'}$ production shall be helpful for confirming whether enough baryon events can be produced and for further testing of the quark model and the nonrelativistic QCD (NRQCD) \cite{Bodwin:1994jh, Petrelli:1997ge}. Here, $\Xi_{bQ'}$ stands for the doubly heavy baryon $\Xi_{bQ'q}$, where $q$ denotes the light $u$, $d$, or $s$ quark, respectively. At present, various methods for the production of doubly heavy baryons have been analyzed in the literature, such as those via the $e^+~e^-$ collisions~\cite{Kiselev:1994pu, Ma:2003zk, Zheng:2015ixa, Jiang:2012jt}, the hadronic collisions~\cite{Berezhnoy:1995fy, Doncheski:1995ye, Baranov:1995rc, Berezhnoy:1998aa, Ma:2003zk, Chang:2006eu, Chang:2007pp, Chang:2009va, Zhang:2011hi, Wang:2012vj, Chen:2014hqa}, the gamma gamma collisions~\cite{Baranov:1995rc, Li:2007vy}, the photoproduction mechanisms~\cite{Baranov:1995rc, Chen:2014frw, Huan-Yu:2017emk}, the heavy ion collisions~\cite{Yao:2018zze, Chen:2018koh}, etc.

It has been confirmed that sizable doubly heavy baryons can be produced at the LHC via the hadronic production mechanisms; their production properties can be simulated by using a dedicated generator GENXICC \cite{Chang:2007pp, Chang:2009va, Wang:2012vj}. In this paper, we shall discuss the production of doubly heavy baryons $\Xi_{bQ'}$ via the top-quark decays at the LHC or its successor [the High Luminosity LHC (HL-LHC)]. Because of the advantage of high luminosity and high energy, the LHC has already become a huge ``top factory''. For example, when the luminosity is up to $10^{34}\sim10^{36}~\rm{cm}^{-2}~\rm{s}^{-1}$, there will be about $10^{8}\sim 10^{10}$ $t\bar t$ pairs produced in one operation year at the LHC \cite{Kidonakis:2004hr, Kuhn:2013zoa}. Thus, the top-quark decays will be a potentially good platform for studying the indirect production mechanism of the doubly heavy baryons and for searching the undetected $\Xi_{bQ'}$. The decay width of the top quark is expected to be dominated by the process $t\to bW^+$. Thus, in the following, we shall study the production of doubly heavy baryons via the process $t\to \Xi_{bQ^{\prime}} +\bar{Q^{\prime}}+ W^+$.

Within the NRQCD framework, the production of $\Xi_{bQ'}$ via top-quark decays can be divided into three steps. The first step is to produce four free heavy particles $b, W^+$ and a heavy quark-antiquark pair $Q^{\prime} \bar {Q^{\prime}}$, which is produced from the intermediate gluon splitting. Since the intermediate gluon, either emitting from the initial or final-state quark, should be hard enough to generate suah a heavy quark-antiquark pair $Q^{\prime}\bar {Q^{\prime}}$, the production process is perturbatively calculable. Then two heavy quarks $b$ and $Q^{\prime}$ shall be coupled into a binding diquark with a corresponding transition probability in the second step. Because of the charge parity, the amplitude for the production of diquark $\langle bQ^{\prime}\rangle$ through $t\rightarrow \langle bQ^{\prime}\rangle +\bar{Q^{\prime}}+ W^+$ is proved to be the same as that of the meson production through $t\rightarrow (b\bar{Q^{\prime}}) +Q^{\prime}+ W^+$ in Refs.~\cite{Jiang:2012jt, Zheng:2015ixa} except for the color factors. The spin state and color state of the intermediate diquark can be $[^3S_1]$ or $[^1S_0]$ and $\bar 3$ or 6 because of the decomposition of $SU(3)_C$ color group. From the symmetry of identical particles in the diquark, there are only two spin and color configurations for $\Xi_{bb}$, $[^3S_1]_{\bar 3}$ and $[^1S_0]_{6}$, while there are four spin and color configurations for $\Xi_{bc}$, $[^3S_1]_{\bar 3}$, $[^3S_1]_{6}$, $[^1S_0]_{\bar 3}$, and $[^1S_0]_{6}$. We shall use $h_{\bar 3}$ and $h_{6}$ to characterize the transition probability of the color antitriplet diquark and the color sextuplet diquark, correspondingly. As will be found later, each of those diquark configurations shall provide sizable contribution to the baryon production; thus, all of those states will be taken into consideration for a sound prediction. The third step is the hadronization from a $\langle bQ^{\prime}\rangle[n]$ diquark state to a doubly heavy baryon $\Xi_{bQ^{\prime}}$, where $n$ stands for the spin and color quantum number of the intermediate diquark. Two methods can be used to deal with the hadronization process: one is the ``direct evolution", which directly set the transition efficiency to be 100\%, and the other is the ``indirect evolution via fragmentation", which can be estimated by using some phenomenological models. It has been found that the direct evolution approach is of high precision and sufficient enough for studying the production of a doubly heavy baryon~\cite{Chen:2014frw}. Thus, in the present paper, we shall directly adopt the direct evolution approach to do the calculation.

In addition to the conventional fixed-order calculations, the production of doubly heavy baryon $\Xi_{bQ^{\prime}}$ from the top-quark decays can also be described by using the fragmentation function approach. The fragmentation function approach resums the large logarithms such as $\ln(M_{\Xi_{bQ^{\prime}}}/E)$ and thus could provide a more reliable prediction in specific kinematic regions, where $E$ is the energy of the corresponding baryon. Such kinds of large logarithms always come from the collinear emissions for high-order calculations or small $p_t$ regions. It is worth mentioning that at leading order in $\alpha_s$, the fragmentation probability $\int_{0}^{1}dzD_{b\rightarrow \Xi_{bQ^{\prime}}}(z,\mu)$ does not evolve with the scale $\mu$~\cite{Chang:1991bp, Chang:1992bb, Braaten:1993jn, Braaten:1996pv}. In the paper, we shall concentrate on the fixed-order calculation with the NRQCD, and shall give a simple discussion on how the fragmentation function approach may change the behaviors of the energy fraction of the produced doubly heavy baryons.

The remaining parts of the paper are organized as follows. In Sec. II, we present the detailed calculation technology for the fixed-order calculation within the framework of NRQCD and the fragmentation function approach. Numerical results and discussions are given in Sec. III. Section IV is reserved for a summary.

\section{Calculation technology}
\label{sec:tech}

\subsection{Fixed-order calculation within the NRQCD framework}
\begin{figure}[htb]
  \centering
  \subfigure[]{
    \includegraphics[scale=0.4]{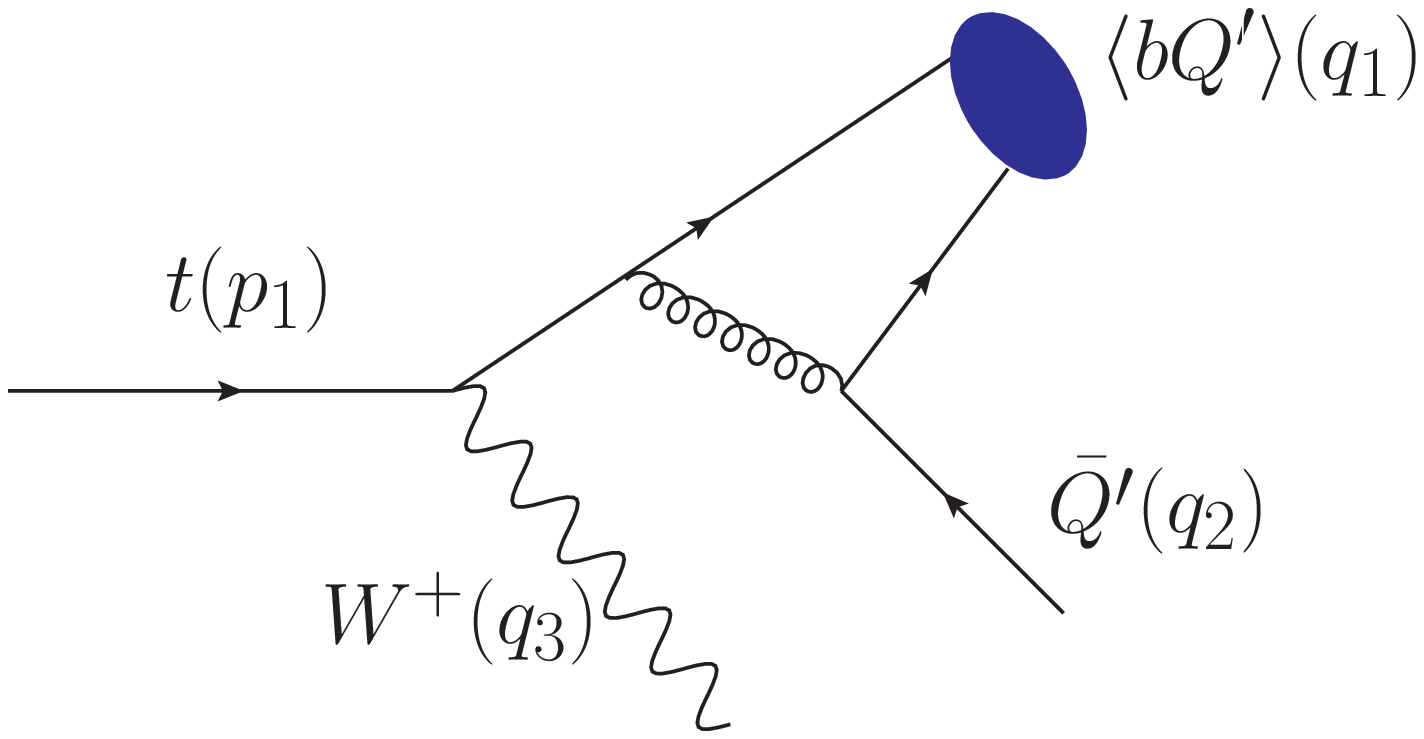}}
  \hspace{0.00in}
  \subfigure[]{
    \includegraphics[scale=0.4]{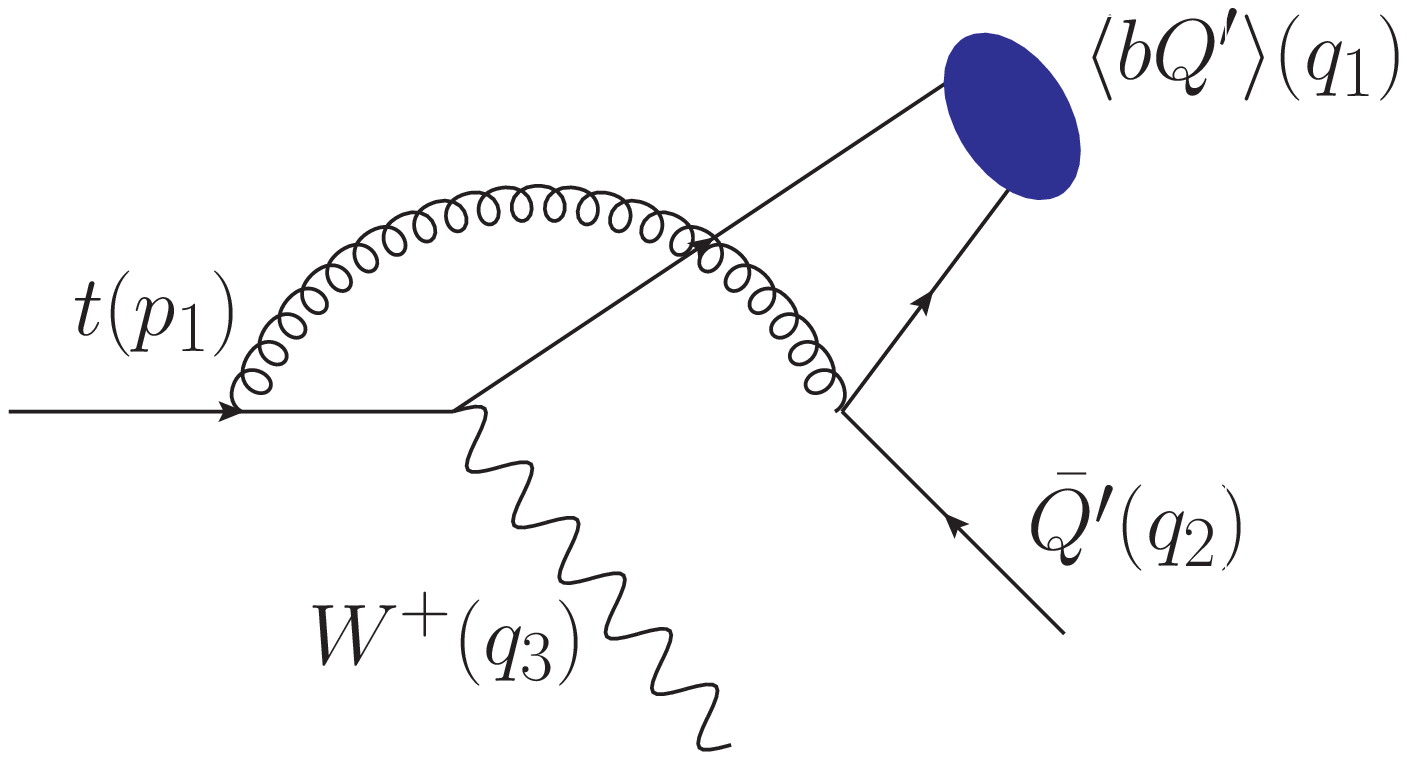}}
  \caption{Typical Feynman diagrams for the process $t\rightarrow \Xi_{bQ^{\prime}}+ \bar {Q^{\prime}} + W^+ $, where $Q^{\prime}$ denotes the heavy $c$ or $b$ quark for the production of $\Xi_{bc}$ or $\Xi_{bb}$ accordingly.}
  \label{diagram1} 
\end{figure}

Typical Feynman diagrams for the process $t\rightarrow \Xi_{bQ^{\prime}}+ \bar{Q^{\prime}} + W^+$ are presented in Fig.~\ref{diagram1}, where $Q^{\prime}$ stands for the heavy $c$ or $b$ quark, respectively. The decay width of this process can be factorized as the following form~\cite{Bodwin:1994jh, Petrelli:1997ge}:
\begin{eqnarray}
\Gamma &&(t \rightarrow \Xi_{bQ^{\prime}}+ \bar {Q^{\prime}} + W^+) =\sum_{n} \hat{\Gamma}(t \rightarrow \langle bQ^{\prime} \rangle[n] + \bar {Q^{\prime}} + W^+) \langle\mathcal O^{H}(n)\rangle.
\end{eqnarray}
Here $[n]$ stands for a series of Fock states with different spin and color quantum numbers for the intermediate diquark state. The nonperturbative matrix element $\langle\mathcal O^{H}(n)\rangle$ is proportional to the transition probability from the perturbative quark pair $b Q^{\prime}$ to the heavy baryon $\Xi_{ bQ^{\prime}}$. The nonperturbative matrix elements are unknown yet, which can be approximated by relating it to the Schr\"{o}dinger wave function at the origin $|\Psi_{bQ^{\prime}}(0)|$ for the S-wave states by assuming the potential of the binding color-antitriplet $\langle bQ^{\prime} \rangle[n]$ state is hydrogenlike. The decay width $\hat{\Gamma}(t \rightarrow \langle bQ^{\prime}\rangle[n] + \bar {Q^{\prime}} + W^+)$ represents the perturbative short-distance coefficients, which can be written as
\begin{eqnarray}
\hat{\Gamma}(t \rightarrow \langle bQ^{\prime}\rangle[n] +\bar {Q^{\prime}} + W^+)= \int \frac{1}{2p_{1}^{0}} \overline{\sum} |\mathcal{M}|^2 d\Phi_3,
\label{width}
\end{eqnarray}
where $\mathcal{M}$ is the hard scattering amplitude. $\overline{\sum}$ means to average over the spin and color of the initial top quark and sum over the colors and spins of all final-state particles. The three-particle phase space $d\Phi_3$ can be represented as
\begin{eqnarray}
d\Phi_3=(2\pi)^4 \delta^4(p_1-\sum_{f=1}^{3} q_f) \prod_{f=1}^{3} \frac{d^3 q_f}{(2\pi)^3 2 q_{f}^{0}}\nonumber.
\end{eqnarray}

After performing the integration over the phase space of this $1 \rightarrow 3$ process, Eq.~(\ref{width}) can be rewritten as
\begin{eqnarray}
d \hat{\Gamma} = \frac{1}{256 \pi^{3} m_{t}^{3}} \overline{\sum} |\mathcal{M}|^2 ds_{12} ds_{23},
\end{eqnarray}
where $m_t$ is the mass of top quark, and the definitions of invariant mass are $s_{12}=(q_{1}+q_{2})^{2}$ and $s_{23}=(q_{2}+q_{3})^{2}$. The VEGAS~\cite{Lepage:1977sw} program is employed to integrate over the invariant mass $s_{12}$ and $s_{23}$. Therefore, not only the total decay width but also the corresponding differential distributions, which are helpful for experimental measurements, can be derived. For useful reference, we shall present the differential distributions in the following discussions, such as $d\Gamma/ds_{12}$, $d\Gamma/ds_{23}$, $d\Gamma/dcos \theta_{13}$, and $d\Gamma/dcos \theta_{12}$, where $\theta_{13}$ is the angle between $\overrightarrow{q_{1}}$ and $\overrightarrow{q_{3}}$ and $\theta_{12}$ is the angle between $\overrightarrow{q_{1}}$ and $\overrightarrow{q_{2}}$.

The hard scattering amplitude $\mathcal M$ for the production of a diquark can be related to the familiar meson production with the action
of $C$ parity, which has been proven in Refs.~\cite{Jiang:2012jt, Zheng:2015ixa}. More explicitly, we can obtain the hard scattering amplitude $\mathcal M$ of the decay channel $t(p_1) \rightarrow \langle bQ^{\prime}\rangle[n](q_1) + \bar Q^{\prime}(q_2) +W^+(q_3)$ from the process $t(p_1) \rightarrow ( b\bar {Q^{\prime}}) [n](q_1) + Q^{\prime}(q_2) +W^+(q_3)$, where there is only a difference between the heavy $Q^{\prime}$ fermion line. Using the charge conjugation matrix $C=-i\gamma^2\gamma^0$, the heavy $Q^{\prime}$ fermion line can be reversed with an additional factor $(-1)^{n+1}$, where $n$ stands for the number of vector vertices and here $n=1$. Hence, we can transform the amplitude of the diquark production to that of the meson production. The square of the hard scattering amplitude
$|\overline{\mathcal{M}}|^{2}=\frac{1}{2 \times 3} \sum |\mathcal{A}_1+\mathcal{A}_2|^{2}$. Here, $\mathcal{A}_1$ ($\mathcal{A}_2$) is the amplitude corresponding to the left (right) diagram of Fig. \ref{diagram1}, which can be written as
\begin{widetext}
\begin{eqnarray}
\mathcal{A}_{1}=i \mathcal{C} \bar{u}_i(q_2, s) \left[\gamma_{\mu} \frac{\Pi_{q_1}[n]}{(q_2+q_{12})^2} \gamma_{\mu} \frac{\slashed{q}_{1}+\slashed{q}_2+m_b}{(q_1+q_2)^2-m_{b}^{2}} \slashed{\varepsilon}(q_3) P_{L}\right] u_j(p_1, s{\prime}), \\
\mathcal{A}_{2}=i \mathcal{C} \bar{u}_i(q_2, s) \left[\gamma_{\mu} \frac{\Pi_{q_1}[n]}{(q_2+q_{12})^2} \slashed{\varepsilon}(q_3) P_{L}  \frac{\slashed{q}_{11}+\slashed{q}_3+m_t}{(q_{11}+q_3)^2-m_{t}^{2}} \gamma_{\mu}\right] u_j(p_1, s{\prime}),
\end{eqnarray}
\end{widetext}
in which the projector
\begin{widetext}
\begin{eqnarray}
\Pi_{q_1}[n] &=& \frac{1}{2\sqrt{M_{bQ^{\prime}}}}\varepsilon[n](\slashed{q}_{1}+ M_{bQ^{\prime}}),
\label{Pieq}
\end{eqnarray}
\end{widetext}
where $\varepsilon[^1S_0]=\gamma_5$ and $\varepsilon[^3S_1]=\slashed{\varepsilon}$ with $\varepsilon^\alpha$ is the polarization vector of the $^3S_1$ diquark state.
$\varepsilon(q_3)$ is the polarization vector of $W^{+}$, $P_L= \frac{1-\gamma_{5}}{2}$. $M_{bQ^{\prime}}=m_{b}+m_{Q^{\prime}}$ is adopted to ensure the gauge invariance, $q_{11}$ and $q_{12}$ are the momenta of those two constituent quarks and take the following forms:
\begin{eqnarray}
q_{11}=\frac{m_{b}}{M_{bQ^{\prime}}}+q ~~\rm{and}~~\it{q}_{\rm{12}}=\it{\frac{m_{Q^{\prime}}}{M_{bQ^{\prime}}}-q},
\end{eqnarray}
where $q$ is the relative momentum between the two constituent quarks of the diquark. On account of the nonrelativistic approximation, $q$ is small enough to be neglected in the amplitude. For the production of $\Xi_{bb}$, we need to time the squared amplitude by an extra overall factor $(2^2/2!)=2$, where the $1/2!$ factor is for the identical particles of the $\langle bb\rangle$ diquark and the $2^2$ factor is because there are two more diagrams coming from the exchange of the two identical
quark lines inside the diquark. The overall factor $\mathcal{C}=g g_{s}^{2} \mathcal{C}_{ij,k}$. Because of the decomposition of the $SU(3)_{C}$ color group $3\bigotimes3=\bar{3} \bigoplus6$, the diquark $\langle bQ^{\prime}\rangle$ can be in either the antitriplet $\bar 3$ or the sextuplet 6 color state. According to Fig. \ref{diagram1}, the color factor $\mathcal{C}_{ij,k}$ can be calculated by
\begin{eqnarray}
\mathcal{C}_{ij,k}=\mathcal{N} \times \sum_{a,m,n} (T^a)_{im} (T^a)_{jn} \times G_{mnk},
\end{eqnarray}
where $i, j, m, n= 1, 2, 3$ are the color indices of the outgoing antiquark $\bar{Q^{\prime}}$, the initial top quark, and the two constituent quarks $b$ and $Q^{\prime}$ of the diquark, correspondingly; $a=1, \ldots, 8$ denotes as the color index for the gluon; $k$ stands for the color index of the diquark $\langle bQ^{\prime}\rangle$; and the normalization constant $\mathcal{N}=\sqrt{1/2}$. For the antitriplet $\bar 3$ color state, the function $G_{mnk}$ is equal to the antisymmetric function $\varepsilon_{mnk}$, which satisfies
\begin{eqnarray}
\varepsilon_{mnk} \varepsilon_{m^{\prime}n^{\prime}k}=\delta_{mm^{\prime}}\delta_{nn^{\prime}}-\delta_{mn^{\prime}}\delta_{nm^{\prime}}.
\end{eqnarray}
The function $G_{mnk}$ stands for the symmetric function $f_{mnk}$ for the sextuplet 6 color state, which satisfies the relation
\begin{eqnarray}
f_{mnk} f_{m^{\prime}n^{\prime}k}=\delta_{mm^{\prime}}\delta_{nn^{\prime}}+\delta_{mn^{\prime}}\delta_{nm^{\prime}}.
\end{eqnarray}
In the squared amplitude, the final color factor $\mathcal{C}^{2}_{ij,k}$ equals $\frac{4}{3}$ for the color antitriplet diquark production and $\frac{2}{3}$ for the color sextuplet diquark production.

According to NRQCD theory, the $\Xi_{bQ^{\prime}}$ baryon can be expanded to a series of Fock states which is accounted by the velocity scaling rule,
\begin{eqnarray}
|\Xi_{bQ^{\prime}}\rangle=c_1(v)|(bQ^{\prime})q\rangle+c_2(v)|(bQ^{\prime})qg\rangle
+c_3(v)|(bQ^{\prime})qgg\rangle+\cdots,
\label{expand}
\end{eqnarray}
where $v$ is the relative velocity of the constituent heavy quarks in the baryon rest frame. For the production of $\Xi_{bb}$, the intermediate diquark states can be either the $\langle bb\rangle[^3S_1]_{\bar{3}}$ state or the $\langle bb\rangle[^1S_0]_6$ state, respectively. And for the case of $\Xi_{bc}$, there are four intermediate diquark states such as $\langle bc\rangle[^3S_1]_{\bar{3}}$, $\langle bc\rangle[^1S_0]_{\bar{3}}$, $\langle bc\rangle[^3S_1]_{6}$, and $\langle bc\rangle[^1S_0]_{6}$. Here, we use $h_{\bar{3}}$ and $h_6$ to present the transition probability of the color antitriplet diquark and the color sextuplet diquark, correspondingly. In the literature, there are two points of view for the contributions from each Fock state. Following the naive NRQCD power counting approach~\footnote{This can be explained by using the ``one-gluon-exchange” interaction picture inside the diquark, i.e. the interaction in the $\bar{3}$ state is attractive, which changes to be repulsive for the case of $6$ state, leading to the fact $h_6 < h_{\bar{3}}$.}, it is generally argued that the $h_6$ should be suppressed by at least $v^2$ to $h_{\bar{3}}$; thus, its contributions can be safely neglected. Another power counting rule indicates that the gluons or light-quarks in the hadron are soft and there is no such $v^2$-power suppressions in the color-sextuplet state~\cite{Ma:2003zk}, and thus, those Fock states in Eq.~(\ref{expand}) are of same importance, i.e., $c_1(v)\sim c_2(v)\sim c_3(v)$. At the present considered pQCD level, the matrix elements are overall parameters, and their uncertainties to the decay width can be conventionally obtained when we have their exact values. For convenience, we will adopt the assumption of the transition probability $h_{\bar 3} \simeq h_6=|\Psi_{bQ^{\prime}}(0)|^2$ \cite{Bagan:1994dy, Petrelli:1997ge} in our discussion.

\subsection{Fragmentation function approach}

In the following, we shall adopt the fragmentation function approach to deal with the process $t\to \Xi_{bQ^{\prime}}+ \bar{Q^{\prime}} + W^+$. At leading order, the energy fraction distribution of the process can be factorized as
\begin{eqnarray}
\frac{d\Gamma}{dz}(t\rightarrow \Xi_{bQ^{\prime}}+W^{+}+X)= \hat\Gamma(t\rightarrow bW^{+})D_{b\rightarrow \Xi_{bQ^{\prime}}}(z,\mu),
\label{fragmentation1}
\end{eqnarray}
where $z$ is the longitudinal momentum fraction of the $\Xi_{bQ^{\prime}}$ relative to the $b$ quark and $z=E_{\Xi_{bQ^{\prime}}}/E^{\rm max}_{\Xi_{bQ^{\prime}}}$, and $\mu$ is the factorization scale. There are large logarithms in the fragmentation function $D_{b\rightarrow \Xi_{bQ^{\prime}}}(z,\mu)$, such as $\ln(M_{\Xi_{bQ^{\prime}}}/E)$, due to collinear gluon emission. Those log terms violate the scaling behavior of the fragmentation function, which, however, can be resummed by using the DGLAP equation~\cite{Altarelli:1977zs, Field:1989uq}, i.e.,
\begin{eqnarray}
\mu\frac{\partial}{\partial\mu}D_{b\rightarrow\Xi_{bQ^{\prime}}}(z,\mu)=
\frac{\alpha_s(\mu_r)}{\pi} \int^{1}_{z} \frac{dy}{y}P_{b\rightarrow b}(z/y)D_{b\rightarrow \Xi_{bQ^{\prime}}}(y,\mu),
\label{dglap}
\end{eqnarray}
and the splitting function $P_{b\rightarrow b}(z$) is
\begin{eqnarray}
P_{b\rightarrow b}(z)=\frac{4}{3}\left(\frac{1+z^2}{1-z}\right)_+.
\label{spliting}
\end{eqnarray}

After performing the integration over the energy fraction $z$ for Eq.~(\ref{fragmentation1}), the fragmentation contribution to the decay rate for the production of $\Xi_{bQ^{\prime}}$ is
\begin{eqnarray}
\Gamma(t\rightarrow \Xi_{bQ^{\prime}}+W^{+}+X)=\hat\Gamma(t\rightarrow bW^{+})\int_{0}^{1} dzD_{b\rightarrow \Xi_{bQ^{\prime}}}(z,\mu).
\label{fragmentation2}
\end{eqnarray}
At leading order in $\alpha_s$, the fragmentation probability $\int_{0}^{1}dzD_{b\rightarrow \Xi_{bQ^{\prime}}}(z,\mu)$ does not evolve with the factorization scale $\mu$ due to the property $\int^1_0dzP_{b\rightarrow b}(z,\mu)=0$~\cite{Braaten:1993jn}. Numerically, the fragmentation probability $\int_{0}^{1}dzD_{b\rightarrow \Xi_{bQ^{\prime}}}(z,\mu)$ can be considered as the branching ratio ${\rm Br}_{t\to\Xi_{bQ^{\prime}}}$ at leading order.

According to the factorization theorem, the heavy baryon fragmentation function (similar to the heavy meson fragmentation function) is independent of the hard processes by which the heavy quark is created. We shall adopt the following fragmentation functions, which are derived by using the $Z^0$-boson decays~\cite{Falk:1993gb, Braaten:1993mp}, to do the numerical calculation, i.e.,
\begin{eqnarray}
D_{b\rightarrow\Xi_{bc}[^1S_0]_{\bar{3}}}(z,\mu=2m_c)&=& \frac{2}{9\pi}\alpha_s(\mu_r)^2\frac{|R_{bc}(0)|^2}{m_c^3}f(z,\frac{m_c}{m_b+m_c}),\nonumber\\
D_{b\rightarrow\Xi_{bc}[^3S_1]_{\bar{3}}}(z,\mu=2m_c)&=& \frac{2}{9\pi}\alpha_s(\mu_r)^2\frac{|R_{bc}(0)|^2}{m_c^3}g(z,\frac{m_c}{m_b+m_c}),\nonumber\\
D_{b\rightarrow\Xi_{bb}[^3S_1]_{\bar{3}}}(z,\mu=2m_b)&=& \frac{4}{9\pi}\alpha_s(\mu_r)^2\frac{|R_{bb}(0)|^2}{m_b^3}F(z),\nonumber\\
D_{b\rightarrow\Xi_{bb}[^1S_0]_{6}}(z,\mu=2m_b)&=& \frac{2}{27\pi}\alpha_s(\mu_r)^2\frac{|R_{bb}(0)|^2}{m_b^3}G(z),
\end{eqnarray}
where
\begin{eqnarray}
|R_{bQ^{\prime}}(0)|^2&=&4\pi|\Psi_{bQ^{\prime}}(0)|^2, \nonumber\\
f(z,r)&=&\frac{rz(1-z)^2}{12[1-(1-r)z]^6}\left[6-18(1-2 r)z+(21-74r+68r^2)z^2-\right.\nonumber\\
 && \left.2(1-r)(6-19r+18r^2)z^3+3(1-r)^2(1-2r+2r^2)z^4\right],\nonumber
  \\
g(z,r)&=&\frac{rz(1-z)^2}{4[1-(1-r)z]^6}\left[2-2(3-2r)z+3(3-2r+4r^2)z ^2-\right.\nonumber\\
 && \left.2(1-r)(4-r+2r^2)z^3+(1-r)^2(3-2r+2r^2)z^4\right],\nonumber\\
 F(z)&=&\frac{z(1-z)^2}{(2-z)^6}\left(16-32z+72z^2-32z^3+5z^4\right),\nonumber\\
 G(z)&=&\frac{z(1-z)^2}{(2-z)^6}\left(48+8z^2-8z^3+3z^4\right).
\end{eqnarray}

\section{Numerical results}

To do the numerical calculation, the input parameters are taken as \cite{Baranov:1995rc, Patrignani:2016xqp}
\begin{eqnarray}
&&m_c=1.8~\rm{GeV},~~~~\it{m_b}=\rm 5.1~{GeV},\nonumber\\
&&M_{\Xi_{bc}}=6.9~\rm{GeV},~~~\it{M_{\rm{\Xi}_{\it{bb}}}}=\rm 10.2~{GeV},\nonumber\\
&&|\Psi_{bc}(0)|^2=0.065~\rm{GeV^3},~|\Psi_{\it{bb}}(0)|^2=\rm 0.152~{GeV^3},\nonumber\\
&&m_t=173.0~\rm{GeV},~~~\it{m_W}=\rm 80.385~{GeV}, \nonumber\\
&&G_{F}=1.1663787 \times 10^{-5}~\rm{GeV^{-2}},~~~\it{g}=\rm 2\sqrt{2}\it m_W \sqrt{G_F/\rm\sqrt{2}},
\end{eqnarray}
where the first six parameters are the same as that in Ref.~\cite{Baranov:1995rc}, $m_c$ and $m_b$ are the constituent quark mass, which is used to build the mass of the corresponding baryon. $|\Psi_{bc}(0)|$ and $|\Psi_{bb}(0)|$ are the Schr\"{o}dinger wave functions at the origin, which can be derived from the potential model, and in our calculation, we adopt the ones evaluated by the power-law potential model~\cite{Bagan:1994dy}. The remaining adoption parameters come from Particle Data Group~\cite{Patrignani:2016xqp}.

The renormalization scale $\mu_r$ for the production of $\Xi_{bc}$ ($\Xi_{bb}$) is set to be $2 m_c$ ($2 m_b$), the same as the factorization scale. With the reference point $\alpha_s(m_{Z})=0.1181$ \cite{Patrignani:2016xqp}, the strong running coupling $\alpha_s(2m_{b})=0.178$ and $\alpha_s(2m_{c})=0.239$ can be obtained from the solution of the five-loop renormalization group equation~\cite{Baikov:2016tgj, Herzog:2017ohr}. To predict the events of the produced $\Xi_{bc}$ and $\Xi_{bb}$ baryons, the total decay width of the top quark is needed. The decay width for the largest decay channel $t \rightarrow bW^{+}$ is 1.49~GeV, which can be considered as the total decay width of the top quark.

\subsection{Basic results}

Based on the input parameters mentioned before, the fixed-order decay widths for all considered spin and color configurations through the process $t\rightarrow \Xi_{bQ^{\prime}}+ \bar {Q^{\prime}}+W^{+}$ are
\begin{eqnarray}
&&\Gamma_{t\rightarrow \Xi_{bc}[^1 S_0]_{\bar{3}}}=0.0962 \;\rm{MeV},\nonumber \\
&&\Gamma_{t\rightarrow \Xi_{bc}[^1 S_0]_6}=0.0481 \rm{MeV}, \;\nonumber\\
&&\Gamma_{t\rightarrow \Xi_{bc}[^3 S_1]_{\bar{3}}}=0.1276 \;\rm{MeV}, \nonumber\\
&&\Gamma_{t\rightarrow \Xi_{bc}[^3 S_1]_6}= 0.0638 \rm{MeV},\; \nonumber\\
&&\Gamma_{t\rightarrow \Xi_{bb}[^1 S_0]_6}=0.00477 \rm{MeV},\; \nonumber\\
&&\Gamma_{t\rightarrow \Xi_{bb}[^3 S_1]_{\bar{3}}}=0.00937\, \rm{MeV},
\label{width11}
\end{eqnarray}
which indicate that
\begin{itemize}
  \item The total decay width for the production of $\Xi_{bb}$ is about 1 order of magnitude smaller than that of $\Xi_{bc}$. The main reason is that the mass of the $b$ quark is about three times larger than that of the $c$ quark, leading to a phase space suppression of $\Gamma_{t \rightarrow \Xi_{bb}}$.

  \item For the production of $\Xi_{bc}$ and $\Xi_{bb}$, the biggest decay width is from the spin and color state $[^{3} S _1]_{\bar {3}}$.

  \item There are four spin and color states for the $\Xi_{bc}$ production. If the transition probability of the color antitriplet diquark $\langle bc\rangle_{\bar{3}}$ and color sextuplet diquark $\langle bc\rangle_6$ are considered the same, i.e., $h_{6} \simeq h_{\bar 3}$, the decay width in the color antitriplet state shall be about two times of that of the color sextuplet. The ratio among the decay widths for the four states of $\Xi_{bc}$ is $[^{3} S _1]_{\bar {3}}:[^{3} S _1]_{6}:[^{1} S _0]_{\bar {3}}:[^{1} S _0]_{6}=1:0.50:0.75:0.38$. In other words, compared to the total decay width $\Gamma_{t \rightarrow \Xi_{bc}}$, the proportions of those four spin and color states are 38\%, 19\%, 29\%, and 14\%, accordingly. All of them have significant contributions, so they need to be considered in the numerical calculation for a sound analysis.

  \item There are only two spin and color states, $\langle bb\rangle [^{3} S _1]_{\bar {3}}$ and $\langle bb\rangle [^{1} S _0]_{6}$, for the production of $\Xi_{bb}$ due to the symmetry of identical particles in the diquark. The ratio between these two states is $[^{3} S _1]_{\bar {3}}:[^{1} S _0]_{6}=1:0.51$. Similar to the case of $\Xi_{bc}$, both of them have sizable contributions to the decay width $\Gamma_{t\rightarrow\Xi_{bb}}$.
\end{itemize}

\begin{table}[htb]
\begin{center}
\caption{The branching ratio ${\rm Br}_{t\rightarrow\Xi_{bQ^{\prime}}}$ under the fixed-order and fragmentation function approaches. The last column shows the ratio of corresponding results under these two approaches.}  \vspace{0.5cm}
\begin{tabular}{cccccc}
\hline
~~~${\rm Br}_{t\rightarrow\Xi_{bQ^{\prime}}}$~~~ &~~~Fixed-order~~~& ~~~Fragmentation function~~~ & ~~~Ratio~~~ \\
\hline
${\rm Br}_{t\rightarrow \Xi_{bc}[^1 S_0]_{\bar{3}}}$ & $6.46 \times 10^{-5}$ & $7.00 \times 10^{-5}$ & 92\%\\
${\rm Br}_{t\rightarrow \Xi_{bc}[^1 S_0]_6}$ & $3.23 \times 10^{-5}$& $3.50 \times 10^{-5}$ & 92\%\\
${\rm Br}_{t\rightarrow \Xi_{bc}[^3 S_1]_{\bar{3}}}$ & $8.56 \times 10^{-5} $& $9.43 \times 10^{-5} $& 91\%\\
${\rm Br}_{t\rightarrow \Xi_{bc}[^3 S_1]_6}$ & $4.28 \times 10^{-5}$ & $4.72 \times 10^{-5} $& 91\%\\
${\rm Br}_{t\rightarrow \Xi_{bb}[^1 S_0]_6}$ & $3.20 \times 10^{-6} $& $3.88 \times 10^{-6} $& 82\% \\
${\rm Br}_{t\rightarrow \Xi_{bb}[^3 S_1]_{\bar{3}}}$ & $6.29 \times 10^{-6}$ & $8.00 \times 10^{-6} $& 79\%\\
\hline
\end{tabular}
\label{ratio12}
\end{center}
\end{table}

The corresponding branching ratios ${\rm Br}_{t\rightarrow\Xi_{bQ^{\prime}}}$ are provided in Table~\ref{ratio12}, in which the results by using the fragmentation function approach are also presented. Table~\ref{ratio12} shows that the results under those two approaches, the fixed-order calculation and fragmentation function approaches, are consistent with each other for the production of $\Xi_{bc}$. To be specific, the ratios for the branching ratio $Br_{t\rightarrow\Xi_{bc}}$ through the fixed-order calculation and fragmentation function approach can be up to $92\%$ for the $[^1 S_0]_{\bar{3}/6}$ state and $91\%$ for the $[^3 S_1]_{\bar{3}/6}$ state. For the case of $\Xi_{bb}$, the ratios change to $82\%$ for the $[^1 S_0]_{6}$ state and $79\%$ for the $[^3 S_1]_{\bar{3}}$ state. Roughly, such a difference can be explained by the fact that the mass of the $b$ quark is heavier than the $c$ quark, leading to fewer $\Xi_{bb}$ events going along with the direction of the $\bar{b}$ quarks. Such an explanation is consistent with that one will find in Fig.~\ref{cos}(b). From Fig.~\ref{cos}(b),
a smaller arising trend for the angular distribution between the baryon $\Xi_{bb}$ and $\bar{b}$ quark can be seen compared to that between the baryon $\Xi_{bc}$ and $\bar{c}$ quark. We also see the peaks of $\Xi_{bb}$ curves are much slower than that of $\Xi_{bc}$ curves.

Table~\ref{ratio12} also shows that the branching ratio ${\rm Br}_{t\rightarrow\Xi_{bQ^{\prime}}}$ through the top-quark decays at the LHC is large enough to be detected. Considering that at the LHC running with a high luminosity $\mathcal{L}$ =$10^{34-36}~\rm cm^{-2} s^{-1}$, about $N_t=10^{8-10} ~t\bar{t}$ pair~\cite{Kidonakis:2004hr, Kuhn:2013zoa} will be produced in one operation year, so the produced $\Xi_{bQ^{\prime}}$ events per year could be roughly estimated by $N_{\Xi_{bQ^{\prime}}}=N_{t} Br_{t\rightarrow\Xi_{bQ^{\prime}}}$:

\begin{itemize}
  \item About $2.25 \times 10^{4-6}$ $\Xi_{bc}$ events/yr will be produced via the top-quark decays at the LHC. The largest proportion comes from the $[^{3} S _1]_{\bar {3}}$ state, which is about $8.56 \times 10^{3-5}$ events/yr. The events per year for the $[^{1} S _0]_{\bar {3}}$, $[^{1} S _0]_{6}$, and $[^{3} S _1]_{6}$ states are about $6.46 \times 10^{3-5}$, $3.23 \times 10^{3-5}$, and $4.28\times 10^{3-5}$, accordingly.

  \item About $9.49 \times 10^{2-4}$ $\Xi_{bb}$ events/yr will be produced via the top-quark decays at the LHC, in which $6.29 \times 10^{2-4}$ events/yr come from the $[^3 S_1]_{\bar{3}}$ state and $3.20 \times 10^{2-4}$ events/yr come from the $[^1 S_0]_{6}$ state.
\end{itemize}

\subsection{Differential decay widths}

\begin{figure}[htb]
  \centering
  \subfigure[]{
    \includegraphics[width=0.47\textwidth]{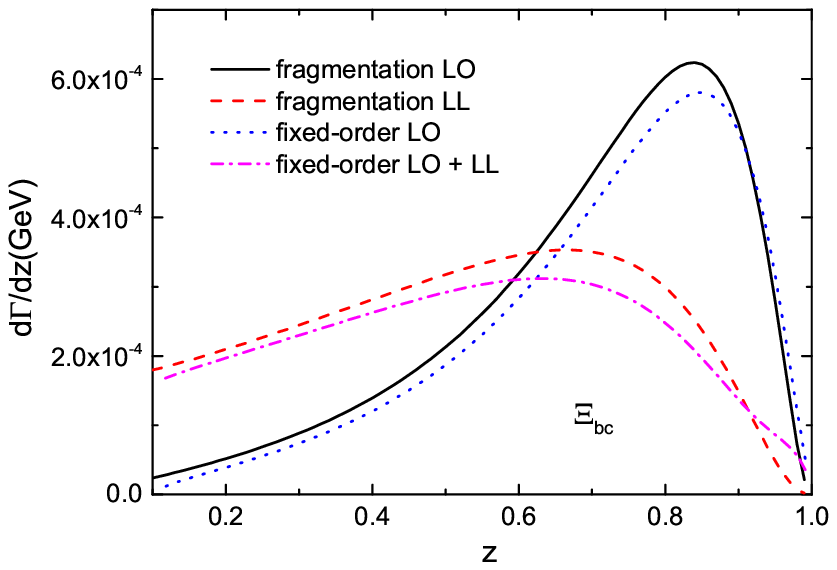}}
  \hspace{0.00in}
  \subfigure[]{
    \includegraphics[width=0.47\textwidth]{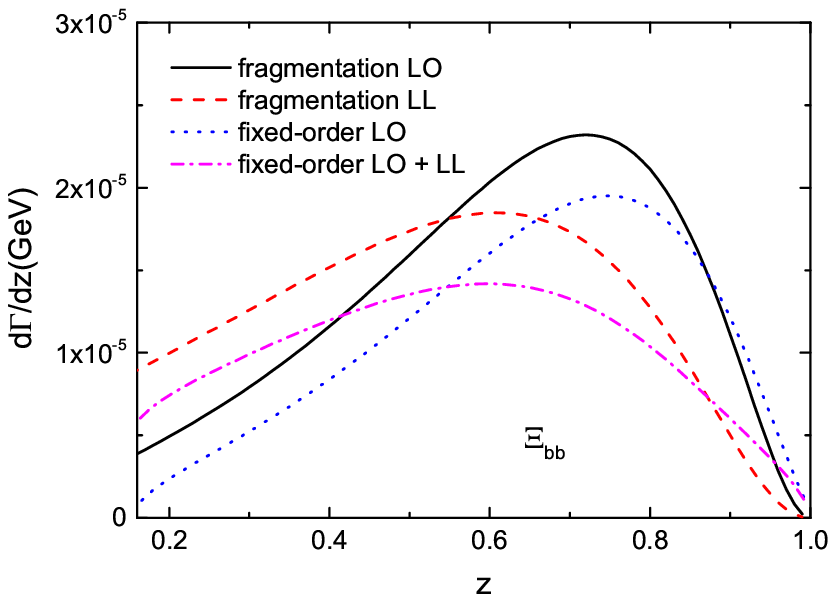}}
  \caption{The differential decay widths $d\Gamma/dz$ for the production of $\Xi_{bc}$ (a) and $\Xi_{bb}$ (b), where $z$ is the energy fraction of the corresponding baryon. The solid black line represents the result obtained by the leading-order fragmentation function approach, the dashed red line denotes that obtained by the leading-logarithm fragmentation function approach, the dotted blue line stands for the fixed-order calculation, and the dash-dotted magenta line is the fixed-order calculation matching with the leading-logarithm fragmentation function approach. All the intermediate diquark states’ contributions have been summed up to obtain the total energy fraction distribution for $t \rightarrow\Xi_{bQ^{\prime}}+W^++X$.}
  \label{dgdz}
\end{figure}

\begin{figure}[htb]
  \centering
  \subfigure[]{
    \includegraphics[width=0.47\textwidth]{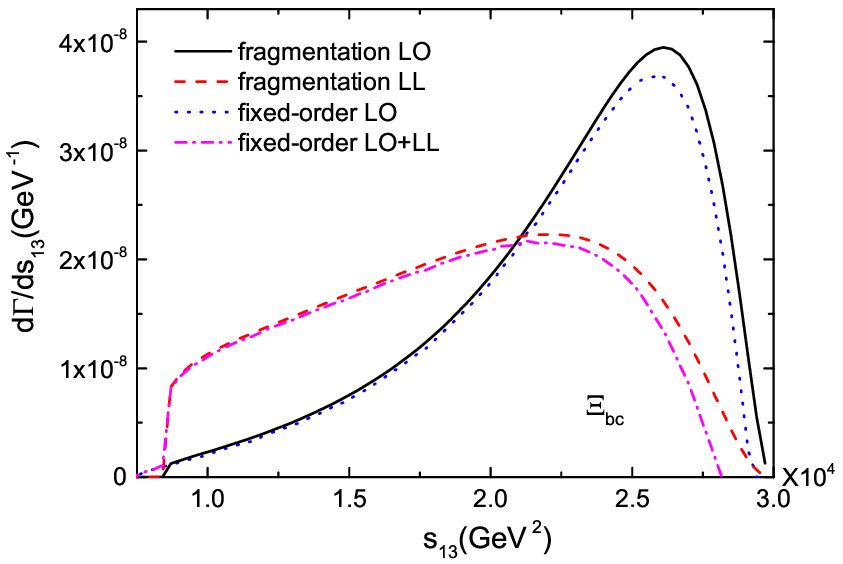}}
  \hspace{0.00in}
  \subfigure[]{
    \includegraphics[width=0.47\textwidth]{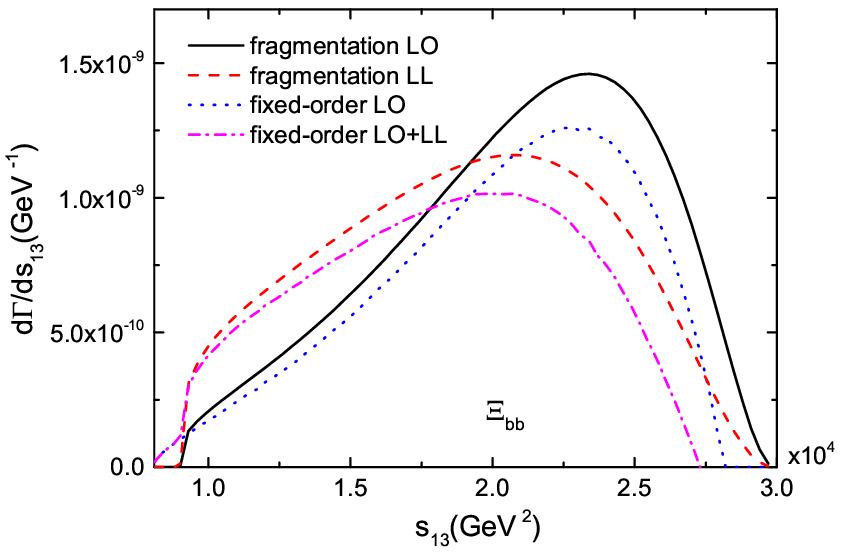}}
  \caption{The invariant mass differential decay width $d\Gamma/ds_{13}$ for the production of $\Xi_{bc}$ (a) and $\Xi_{bb}$ (b). The solid black line represents the result obtained by the leading-order fragmentation function approach, the dashed red line denotes that obtained by the leading-logarithm fragmentation function approach, the dotted blue line stands for the fixed-order calculation, and the dash-dotted magenta line is the fixed-order calculation matching with the leading-logarithm fragmentation function approach. All the intermediate diquark states’ contributions have been summed up to obtain the total energy fraction distribution for $t \rightarrow\Xi_{bQ^{\prime}}+W^++X$.}
  \label{dgds13}
\end{figure}

To make a clear analysis about the distribution that is helpful to the experiments detection, we present the energy fraction distributions for the production of $\Xi_{bc}$ (a) and $\Xi_{bb}$ (b) in Fig.~\ref{dgdz}, in which both the results for the fixed-order calculation and fragmentation function approach are presented. In doing the calculation, all the parameters are set to be their central values. The fragmentation function approach is at the leading-logarithm (LL) accuracy, the factorization scale of which has been evolved to the order of $m_t-m_W$. The contribution from the large logarithms can be extracted by subtracting the fragmentation leading-order (LO) contribution from the fragmentation LL. After matching the fixed-order results with large logarithms contribution from the LL, an overall fixed-order LO+LL result can be given. In drawing the curves, the contributions from different diquark spin and color configurations have been summed up. Figure~\ref{dgdz} shows that the behavior of the energy fraction of the produced doubly heavy baryon under the fixed-order calculation is close in shape in comparison to that of the leading-order fragmentation function approach, being without resumming the large logs. After doing the resummation at the LL level, the distributions of doubly heavy baryons $\Xi_{bc}$ and $\Xi_{bb}$ in the low-energy fraction (small $z$ region) estimated by the LL fragmentation function approach are greater than those of the fixed-order calculation.

Meanwhile, the invariant mass differential decay width $d\Gamma/ds_{13}$ is displayed in Fig.~3 for the production of $\Xi_{bQ'}$ by these two approaches and the combined one. Figure~3 shows that the behavior of the invariant mass distribution is analogous to that of the $z$ distribution. In the fragmentation function approach, the considered process for the production of $\Xi_{bQ^{\prime}}$ is $t(p_1) \rightarrow bW^+\rightarrow\Xi_{bQ^{\prime}}(q_1)+W^+(q_3)+X$ produced heavy flavor baryon $\Xi_{bQ'}$ and $W^+$ boson, which are back-to-back with an angular separation by $\theta=\pi$ at the LL level. For the angular distribution obtained by the fragmentation function approach, $d\Gamma/dcos\theta_{13}$ is represented by a delta function, $\delta(\theta-\pi)$. And after the resummation with the DGLAP equation, it does not change the direction of the momentum, and the angular distribution obtained by the fixed-order calculation is not changed after being revised by the fragmentation function approach. The future experimental data may be helpful to test the result of theoretical predictions.

\begin{figure}[htb]
  \centering
  \subfigure[]{
    \includegraphics[width=0.47\textwidth]{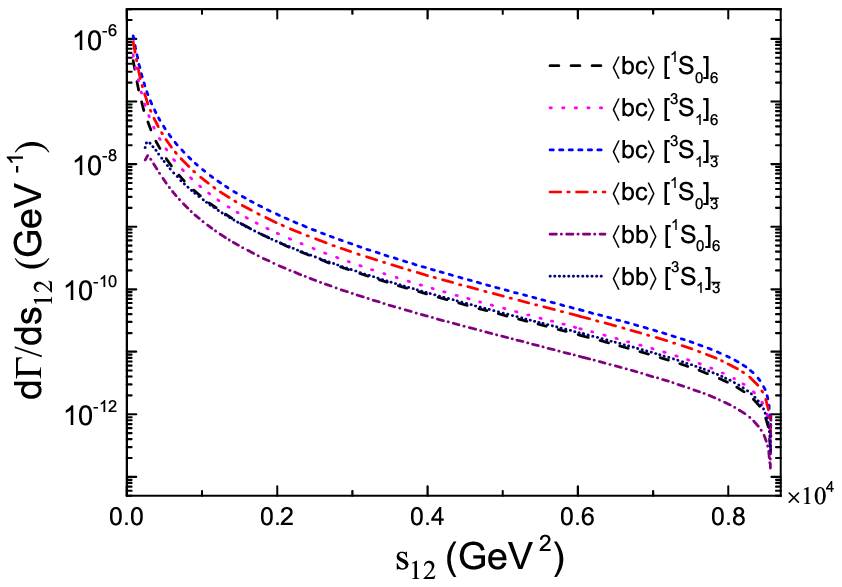}}
  \hspace{0.00in}
  \subfigure[]{
    \includegraphics[width=0.47\textwidth]{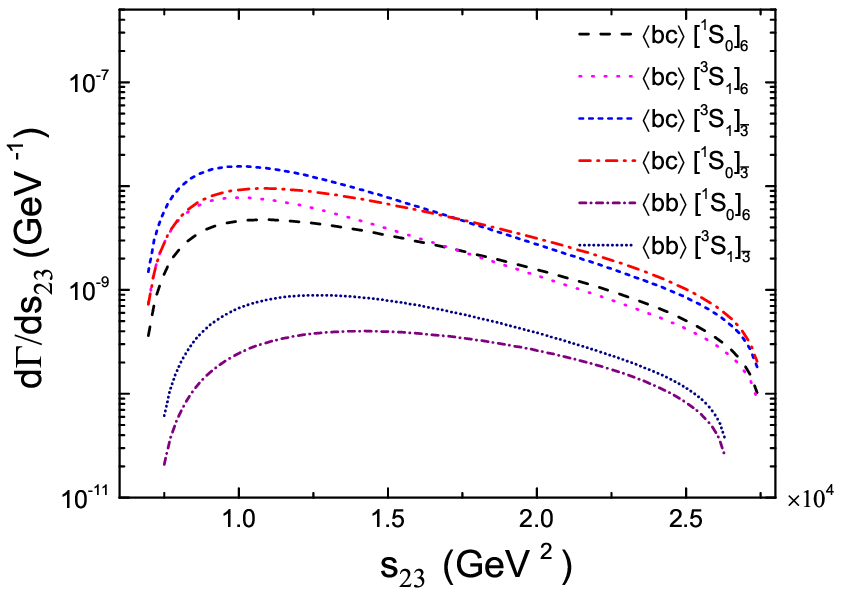}}
  \caption{The differential decay widths $d\Gamma/ds_{12}$ (a) and $d\Gamma/ds_{23}$ (b) for the process $t(p_1) \rightarrow \Xi_{bQ^{\prime}} (q_1)+ \bar {Q^{\prime}} (q_2) + W^{+} (q_3)$. The dashed black, dotted magenta, short dashed blue, dash-dotted red, short dash-dotted purple, and short dotted navy lines represent for the decay width for the production of $\Xi_{bQ^{\prime}}$ in Fock states: $\langle bc \rangle [^{1}S_{0}]_{6}$, $\langle bc \rangle [^{3}S_{1}]_{6}$, $\langle bc \rangle [^{3}S_{1}]_{\bar{3}}$, $\langle bc \rangle [^{1}S_{0}]_{\bar{3}}$, $\langle bb \rangle [^{1}S_{0}]_{6}$, and $\langle bb \rangle [^{3}S_{1}]_{\bar{3}}$, respectively.}
  \label{s1s2}
\end{figure}

\begin{figure}[htb]
  \centering
  \subfigure[]{
    \includegraphics[width=0.47\textwidth]{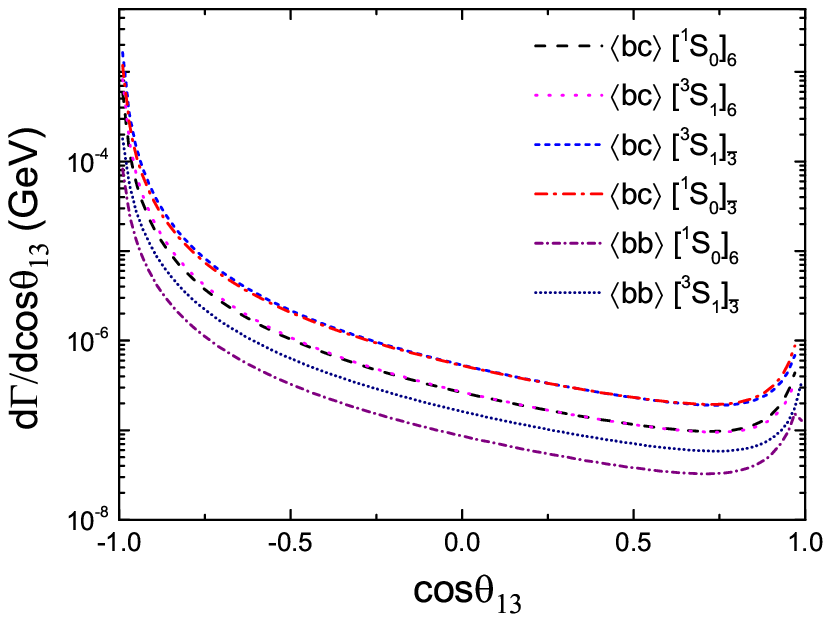}}
  \hspace{0.00in}
  \subfigure[]{
    \includegraphics[width=0.47\textwidth]{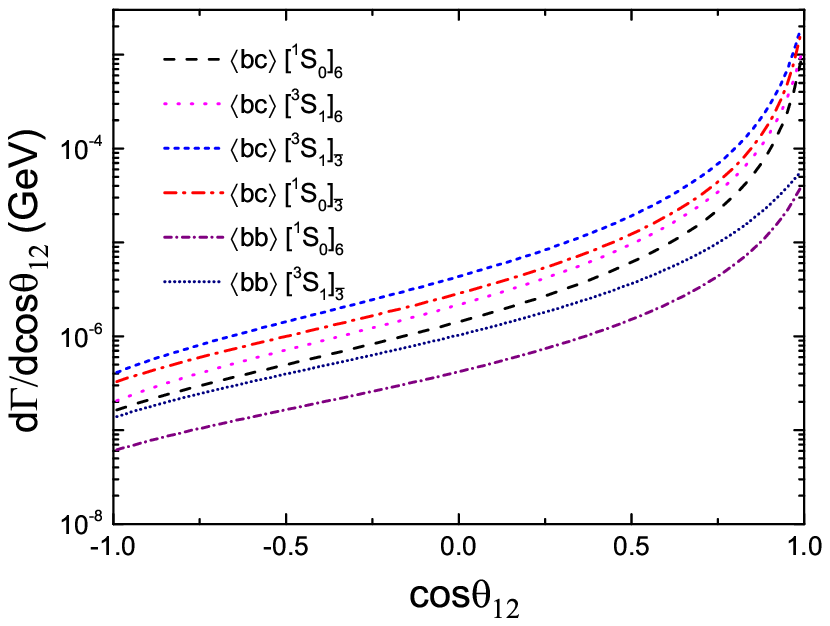}}
  \caption{The differential decay widths  $d\Gamma/dcos\theta_{13}$ (a) and $d\Gamma/dcos\theta_{12}$ (b) for the process $t(p_1) \rightarrow \Xi_{bQ^{\prime}} (q_1)+ \bar {Q^{\prime}} (q_2) + W^{+} (q_3)$. The dashed black, dotted magenta, short dashed blue, dash-dotted red, short dash-dotted purple, and short dotted navy lines represent for the decay width for the production of $\Xi_{bQ^{\prime}}$ in Fock states: $\langle bc \rangle [^{1}S_{0}]_{6}$, $\langle bc \rangle [^{3}S_{1}]_{6}$, $\langle bc \rangle [^{3}S_{1}]_{\bar{3}}$, $\langle bc \rangle [^{1}S_{0}]_{\bar{3}}$, $\langle bb \rangle [^{1}S_{0}]_{6}$, and $\langle bb \rangle [^{3}S_{1}]_{\bar{3}}$, respectively.}
  \label{cos}
\end{figure}

More characteristics of the process $t(p_1) \rightarrow \Xi_{bQ^{\prime}} (q_1)+ \bar {Q^{\prime}} (q_2) + W^{+} (q_3)$ are obtained by the fixed-order calculation, such as the differential decay widths $d\Gamma/ds_{12}$, $d\Gamma/ds_{23}$, $d\Gamma/dcos\theta_{12}$ and $d\Gamma/dcos\theta_{13}$ which are shown in Figs.~\ref{s1s2} and \ref{cos}. The kinematics parameters $s_{12}$, $s_{23}$, and $\cos\theta$ are defined in Sec. \ref{sec:tech}.

Figure~\ref{s1s2}(a) shows that the differential decay width monotonously decreases  with the increment of $s_{12}$. For smaller and smaller $s_{12}$, the $\Xi_{bQ'}$ baryon shall move closer to the direction of the heavy quark $\bar{Q^{\prime}}$, leading to a much larger decay width. For the extreme condition, the $\Xi_{bQ^{\prime}}$ baryon and the heavy quark $\bar{Q^{\prime}}$ shall move in the same direction, both of which shall move back to back with the $W^{+}$ boson in the rest frame of the top quark. In Fig.~\ref{s1s2}(b), all the curves are relatively flatter than those in Fig.~\ref{s1s2}(a), which first increase and then decrease with the increment of $s_{23}$ and have maximum values in the small $s_{23}$ region.

Figure~\ref{cos}(a) shows that when the outgoing $\Xi_{bQ^{\prime}}$ and $W^{+}$ boson move back to back, i.e., $\theta_{13}=180^{\rm{o}}$, the differential decay width $d\Gamma/dcos{\theta_{13}}$ can achieve its largest value, which is due to the fact that the $W^{+}$ boson is the heaviest among these three outgoing particles. Figure~\ref{cos}(b) shows that when the outgoing $\Xi_{bQ^{\prime}}$ and antiquark $\bar{Q^{\prime}}$ move in the same direction, i.e., $\theta_{12}=0^{\rm{o}}$, the differential decay width $d\Gamma/dcos\theta_{12}$ can achieve its largest value, which agrees with the result of Fig.~\ref{s1s2}(a).

\subsection{Theoretical uncertainties}

In this subsection, theoretical uncertainties for the production of $\Xi_{bQ^{\prime}}$ through top-quark decays under the fixed-order calculation shall be discussed.

\begin{table}[htb]
\begin{center}
\caption{The theoretical uncertainty for the production of $\Xi_{bc}$ via top-quark decays by varying $m_c=1.8 \pm 0.3~\rm{GeV}$ with $m_b=5.1~\rm{GeV}$ and $m_t=173.0~\rm{GeV}$ at their central values.} \vspace{0.5cm}
\begin{tabular}{cccccc}
  \hline
  ~~~$m_c$ (GeV)~~~ & ~~~1.5~~~ & ~~~1.65~~~ & ~~~1.8~~~ & ~~~1.95~~~ & ~~~2.1~~~ \\
  \hline
  $\Gamma_{\Xi_{bc}[^3 S_1]_{\bar{3}}}$(MeV) & 0.2360 & 0.1711 & 0.1276 & 0.0973 & 0.0757 \\
  $\Gamma_{\Xi_{bc}[^3 S_1]_{6}}      $(MeV) & 0.1180 & 0.0856 & 0.0638 & 0.0487 & 0.0379 \\
  $\Gamma_{\Xi_{bc}[^1 S_0]_{\bar{3}}}$(MeV) & 0.1683 & 0.1257 & 0.0962 & 0.0752 & 0.0599 \\
  $\Gamma_{\Xi_{bc}[^1 S_0]_{6}}$(MeV)       & 0.0842 & 0.0629 & 0.0481 & 0.0376 & 0.0299 \\
  \hline
\end{tabular}
\label{mcun}
\end{center}
\end{table}

\begin{table}[htb]
\begin{center}
\caption{The theoretical uncertainty for the production of $\Xi_{bc}$ and $\Xi_{bb}$ via top-quark decays by varying $m_b=5.1 \pm 0.4~\rm{GeV}$ with $m_c=1.8~\rm{GeV}$ and $m_t=173.0~\rm{GeV}$ at their central values.} \vspace{0.5cm}
\begin{tabular}{cccccc}
  \hline
  ~~~$m_b$ (GeV)~~~ & ~~~4.7~~~ & ~~~4.9~~~ & ~~~5.1~~~ & ~~~5.3~~~ & ~~~5.5~~~ \\
  \hline
  $\Gamma_{\Xi_{bc}[^3 S_1]_{\bar{3}}}$(MeV) & 0.1253 & 0.1265 & 0.1276 & 0.1286 & 0.1296 \\
  $\Gamma_{\Xi_{bc}[^3 S_1]_{6}}      $(MeV) & 0.0627 & 0.0632 & 0.0638 & 0.0643 & 0.0648 \\
  $\Gamma_{\Xi_{bc}[^1 S_0]_{\bar{3}}}$(MeV) & 0.0967 & 0.0965 & 0.0962 & 0.0960 & 0.0957 \\
  $\Gamma_{\Xi_{bc}[^1 S_0]_{6}}$(MeV)       & 0.0484 & 0.0482 & 0.0481 & 0.0480 & 0.0479 \\
  $\Gamma_{\Xi_{bb}[^3 S_1]_{\bar{3}}}$(MeV) & 0.01233 & 0.01072 & 0.00937 & 0.00822 & 0.00724 \\
  $\Gamma_{\Xi_{bb}[^1 S_0]_{6}}$(MeV)       & 0.00624 & 0.00544 & 0.00477 & 0.00420 & 0.00371 \\
  \hline
\end{tabular}
\label{mbun}
\end{center}
\end{table}

\begin{table}[htb]
\begin{center}
\caption{The theoretical uncertainty for the production of $\Xi_{bc}$ and $\Xi_{bb}$ via top-quark decays by varying $m_t=173.0 \pm 0.4~\rm{GeV}$ with $m_c=1.8~\rm{GeV}$ and $m_b=5.1~\rm{GeV}$ at their central values.} \vspace{0.5cm}
\begin{tabular}{cccccc}
  \hline
  ~~~$m_t$ (GeV)~~~ & ~~~172.6~~~ & ~~~173.0~~~ & ~~~173.4~~~  \\
  \hline
  $\Gamma_{\Xi_{bc}[^3 S_1]_{\bar{3}}}$(MeV) & 0.1265 & 0.1276 & 0.1287  \\
  $\Gamma_{\Xi_{bc}[^3 S_1]_{6}}      $(MeV)  & 0.0632 & 0.0638 & 0.0643 \\
  $\Gamma_{\Xi_{bc}[^1 S_0]_{\bar{3}}}$(MeV)  & 0.0954 & 0.0962 & 0.0971  \\
  $\Gamma_{\Xi_{bc}[^1 S_0]_{6}}$(MeV)       & 0.0477 & 0.0481 & 0.0485  \\
  $\Gamma_{\Xi_{bb}[^3 S_1]_{\bar{3}}}$(MeV)  & 0.00928 & 0.00937 & 0.00946  \\
  $\Gamma_{\Xi_{bb}[^1 S_0]_{6}}$(MeV)       & 0.00473  & 0.00477 & 0.00481  \\
  \hline
\end{tabular}
\label{mtun}
\end{center}
\end{table}

First, uncertainties from the main error sources, e.g., $m_c$, $m_b$, and $m_t$, are shown in Tables~\ref{mcun},~\ref{mbun}, and \ref{mtun}, respectively. To estimate the uncertainties, we adopt $m_c=1.8\pm0.3~\rm{GeV}$, $m_b=5.1\pm0.4~\rm{GeV}$, and $m_t=173.0\pm0.4~\rm{GeV}$. It is worth mentioning that $m_c$ and $m_b$ are the essential mass uncertainty factors for building the mass of the corresponding baryon $\Xi_{bQ^{\prime}}$. For clarity, when discussing the uncertainty caused by one parameter, the others shall be fixed to their central values. Tables~\ref{mcun},~\ref{mbun}, and \ref{mtun} show the following:
\begin{itemize}
  \item The decay width for the production of the $\Xi_{bc}$ baryon decreases with the increment of $m_c$, which is mainly due to the suppression of phase space. The uncertainty from $m_c$ is relatively larger than those of $m_b$ and $m_t$. Because of the influence of the projector in Eq.~(\ref{Pieq}), the decay width of $\Xi_{bc}[^3 S_1]$ increases with the increment of $m_b$, and the decay width of $\Xi_{bc}[^1 S_0]$ decreases with the increment of $m_b$.
  \item For the production of $\Xi_{bb}$, the decay width decreases with the increment of $m_b$. The same as in the case of $\Xi_{bc}$, its decay width will also increase with the increment of $m_t$. The theoretical uncertainty from $m_b$ is bigger than that from $m_t$.
\end{itemize}

Second, uncertainty comes from the renormalization scale $\mu_r$ presented in Table~\ref{mqun}, in which we use three different renormalization scales $\mu_r=2m_c$, $\rm{M}_{bc}$, and $2m_b$ for the production of $\Xi_{bQ^{\prime}}$ via top-quark decays. And the corresponding running coupling $\alpha_s$ is also added in Table~\ref{mqun}. Obviously, there is a large uncertainty caused by the renormalization scale $\mu_r$. Such scale ambiguity could be suppressed by a higher-order perturbative calculation or proper scale-setting methods such as the newly suggested principle of maximum conformal \cite{Brodsky:2011ta, Brodsky:2012rj, Mojaza:2012mf, Brodsky:2013vpa}, which uses the renormalization group equation to set the optimal behavior of the running coupling at each perturbative order and thus set the optimal value for the renormalization scale.

\begin{table}[htb]
\begin{center}
\caption{The theoretical uncertainty for the production of $\Xi_{bQ^{\prime}}$ via top-quark decays by substituting the renormalization scale $\mu_r=2m_c$, $\rm{M}_{bc}$, or $2m_b$ with the masses of heavy quarks at their central values.}
\begin{tabular}{cccc}
  \hline
  ~~~$\mu_r$~~~ & ~~~$2m_c$~~~ & ~~~$\rm{M}_{bc}$~~~ & ~~~2$m_b$~~~ \\
  $\alpha_s$ & 0.239 &0.196 & 0.178 \\
  \hline
  $\Gamma_{\Xi_{bc}[^3 S_1]_{\bar{3}}}$(MeV) & 0.1276  & 0.0858 & 0.0708  \\
  $\Gamma_{\Xi_{bc}[^3 S_1]_{6}}      $(MeV) & 0.0638 & 0.0429 & 0.0354  \\
  $\Gamma_{\Xi_{bc}[^1 S_0]_{\bar{3}}}$(MeV) & 0.0962 & 0.0647 & 0.0534  \\
  $\Gamma_{\Xi_{bc}[^1 S_0]_{6}}$(MeV)       & 0.0481 & 0.0324 & 0.0267  \\
  $\Gamma_{\Xi_{bb}[^3 S_1]_{\bar{3}}}$(MeV) & 0.01688 & 0.01136 & 0.00937 \\
  $\Gamma_{\Xi_{bb}[^1 S_0]_{6}}$(MeV)       & 0.00860 & 0.00578 & 0.00477 \\
  \hline
\end{tabular}
\label{mqun}
\end{center}
\end{table}

Finally, uncertainty caused by choices of the nonperturbative transition probability is considered. According to NRQCD, $h_6$ for the color sextuplet diquark state may be suppressed by $v^2$ compared to $h_{\bar 3}$ for the color antitriplet diquark state, such as $h_{6}/v^2 \simeq h_{\bar 3}=|\Psi_{bQ^{\prime}}(0)|^2$. If the contribution from the color sextuplet diquark $\langle bQ^{\prime}\rangle_6$ state can be ignored ($h_{6}=0$) and only the color antitriplet diquark $\langle bQ^{\prime}\rangle_{\bar{3}}$ state is taken into consideration ($h_{\bar 3}=|\Psi_{bQ^{\prime}}(0)|^2$) for the production of $\Xi_{bQ^{\prime}}$, there are still $1.50 \times 10^{4-6}$ events of $\Xi_{bc}$ and $6.29 \times 10^{2-4}$ events of $\Xi_{bb}$ produced at the LHC in one operation year. As for the uncertainty of $h_{\bar 3}$, it can be related to the Schr\"{o}dinger wave function at the origin $|\Psi_{bQ^{\prime}}(0)|$ for the S-wave state. And the wave function at the zero is an overall factor, and its uncertainty can be conventionally discussed when we know its exact values; thus, we directly take the wave function at zero to be the one derived from the power-law potential model~\cite{Bagan:1994dy}.

\section{Summary}

In this paper, the indirect production of doubly heavy baryons $\Xi_{bc}$ and $\Xi_{bb}$ via top-quark decays are discussed under two different approaches: the fixed-order calculation and the fragmentation function approach. In our calculation, all the possible spin and color configurations have been taken into consideration, i.e., $\langle bc\rangle[^3 S_1]_{\bar{3}/6}$, $\langle bc\rangle[^1 S_0]_{\bar{3}/6}$, $\langle bb\rangle[^1 S_0]_{6}$ and $\langle bb\rangle[^3 S_1]_{\bar{3}}$. We observe that each spin and color configuration has a sizable contribution to the production of $\Xi_{b Q^{\prime}}$. By summing up all the intermediate diquark states' contributions, we obtain the total decay width for $t \rightarrow\Xi_{bQ^{\prime}}+\bar{Q^{\prime}}+W^+$,
\begin{eqnarray}
&&\Gamma_{t\rightarrow \Xi_{bc}+\bar{c}+W^+}=0.34^{+0.27}_{-0.15} ~\rm{MeV},\nonumber \\
&&\Gamma_{t\rightarrow \Xi_{bb}+\bar{b}+W^+}=0.014^{+0.011}_{-0.005} ~\rm{MeV},\nonumber
\end{eqnarray}
where the uncertainties are squared averages of those from the heavy quark masses ($m_b,~m_c$ and $m_t$), the renormalization scale $\mu_r$, and the nonperturbative transition probability. The decay width for the production of $\Xi_{bc}$ ($\Xi_{bb}$) is sensitive to $m_c$ ($m_b$), which is mainly caused by the change of phase space. Because of the running behavior of $\alpha_s(\mu_r)$, the renormalization scale $\mu_r$ has a significant impact on the decay width $\Gamma_{\Xi_{bQ^{\prime}}}$. Thus, a proper QCD renormalization scale-setting method~\cite{Wu:2013ei, Wu:2014iba} or higher-order perturbative calculation is needed to eliminate this scale ambiguity.

Considering that at the LHC with a high luminosity $\mathcal{L}=10^{34-36} \rm  ~cm^{-2}s^{-1}$, there will be about $2.25\times 10^{4-6}$ $\Xi_{bc}$ events and $9.49\times 10^{2-4}$ $\Xi_{bb}$ events produced in one operation year. This phenomenon will possibly be found at the LHC (HL-LHC) in the future. To form the doubly heavy baryon $\Xi_{b Q^{\prime}q}$, the intermediate diquark $\langle b Q^{\prime} \rangle [n]$ state needs to grab a light $u$, $d$, or $s$ quark. According to the ratio for the production of $\Xi_{b Q^{\prime}u}$, $\Xi_{b Q^{\prime}d}$ and $\Xi_{b Q^{\prime}s}$ is 1:1:0.3 \cite{Sjostrand:2006za}, there will be about 43\% $\langle b c \rangle[n]$ fragmented into $\Xi_{bc}^{+}$, 43\% to $\Xi_{bc}^{0}$, and 14\% to $\Omega_{bc}^{0}$, and the same percentage for $\langle bb \rangle[n]$ fragmented into $\Xi_{bb}^{0}$, $\Xi_{bb}^{-}$, and $\Omega_{bb}^{-}$. Therefore, there are still sizable events of detectable baryons at the LHC or HL-LHC.

As a final remark, at present,  many phenomenological models have been suggested to study the decay properties of the doubly heavy baryons. However, due to the large nonperturbative effects in those decays, such kinds of studies are at the initial stage. An overview of those decays, together with their possibilities of observation, can be found in Refs.\cite{Bediaga:2012py, Bediaga:2018lhg}. Similar to the observation of the $\Xi_{cc}^{++}$ baryon, the $\Xi_{bc}$ baryon could be observed by cascade decays such as $\Xi_{bc}^{+}\to\Xi_{cc}^{++}(\to p K^-\pi^+ \pi^+)\pi^{-}$, and the $\Xi_{bb}$ baryon could be observed via $\Xi^{0}_{bb}\to\Xi_{bc}^+ (\to\Xi^{++}\pi^-) \pi^-$.

\hspace{2cm}

{\bf Acknowledgements}: This work was partially supported by the National Natural Science Foundation of China (Grants No.11375008, No.11647307, and No.11625520). This research was also supported by Conselho Nacional de Desenvolvimento Cient\'{\i}fico e Tecnol\'ogico (CNPq) and Coordena\c{c}\~ao de Aperfei\c{c}oamento de Pessoal de N\'ivel Superior (CAPES).


\begin{thebibliography}{99}

\bibitem{Aaij:2017ueg}
  R.~Aaij {\it et al.} (LHCb Collaboration),
  Observation of the Doubly Charmed Baryon $\Xi_{cc}^{++}$,
  Phys.\ Rev.\ Lett.\  {\bf 119}, 112001 (2017).

\bibitem{GellMann:1964nj}
  M.~Gell-Mann,
  A Schematic Model of Baryons and Mesons,
  Phys.\ Lett.\  {\bf 8}, 214 (1964).

\bibitem{Zweig:1981pd}
  G.~Zweig,
  An SU(3) model for strong interaction symmetry and its breaking, Version 1,
  Report No. CERN-TH-401.

\bibitem{Zweig:1964jf}
  G.~Zweig,
  An SU(3) Model for Strong Interaction Symmetry and its Breaking, version 2,
  Developments in the Quark Theory of Hadrons, edited by D. Lichtenberg and S. Rosen, Vol.1, p. 22.

\bibitem{DeRujula:1975qlm}
  A.~De Rujula, H.~Georgi, and S.~L.~Glashow,
  Hadron Masses in a Gauge Theory,
  Phys.\ Rev.\ D {\bf 12}, 147 (1975).


\bibitem{Bodwin:1994jh}
  G.~T.~Bodwin, E.~Braaten, and G.~P.~Lepage,
  Rigorous QCD analysis of inclusive annihilation and production of heavy quarkonium,
  Phys.\ Rev.\ D {\bf 51}, 1125 (1995); {\bf 55}, 5853(E) (1997)].

\bibitem{Petrelli:1997ge}
  A.~Petrelli, M.~Cacciari, M.~Greco, F.~Maltoni, and M.~L.~Mangano,
  NLO production and decay of quarkonium,
  Nucl.\ Phys.\ B {\bf 514}, 245 (1998).

\bibitem{Kiselev:1994pu}
  V.~V.~Kiselev, A.~K.~Likhoded and M.~V.~Shevlyagin,
  Double charmed baryon production at B factory,
  Phys.\ Lett.\ B {\bf 332}, 411 (1994).

\bibitem{Ma:2003zk}
  J.~P.~Ma and Z.~G.~Si,
  Factorization approach for inclusive production of doubly heavy baryon,
  Phys.\ Lett.\ B {\bf 568}, 135 (2003).

\bibitem{Zheng:2015ixa}
  X.~C.~Zheng, C.~H.~Chang, and Z.~Pan,
  Production of doubly heavy-flavored hadrons at $e^+e^-$ colliders,
  Phys.\ Rev.\ D {\bf 93}, 034019 (2016).

\bibitem{Jiang:2012jt}
  J.~Jiang, X.~G.~Wu, Q.~L.~Liao, X.~C.~Zheng, and Z.~Y.~Fang,
  Doubly heavy baryon production at a high luminosity $e^+ e^-$ collider,
  Phys.\ Rev.\ D {\bf 86}, 054021 (2012).

\bibitem{Berezhnoy:1995fy}
  A.~V.~Berezhnoy, V.~V.~Kiselev and A.~K.~Likhoded,
  Hadronic production of baryons containing two heavy quarks,
  Yad.\ Fiz.\  {\bf 59}, 909 (1996);
  Phys.\ At.\ Nucl.\  {\bf 59}, 870 (1996);
  [hep-ph/9507242].

\bibitem{Doncheski:1995ye}
  M.~A.~Doncheski, J.~Steegborn, and M.~L.~Stong,
  Fragmentation production of doubly heavy baryons,
  Phys.\ Rev.\ D {\bf 53}, 1247 (1996).

\bibitem{Baranov:1995rc}
  S.~P.~Baranov,
  On the production of doubly flavored baryons in $pp$, $e p$ and $\gamma \gamma$ collisions,
  Phys.\ Rev.\ D {\bf 54}, 3228 (1996).

\bibitem{Berezhnoy:1998aa}
  A.~V.~Berezhnoy, V.~V.~Kiselev, A.~K.~Likhoded and A.~I.~Onishchenko,
  Doubly charmed baryon production in hadronic experiments,
  Phys.\ Rev.\ D {\bf 57}, 4385 (1998).

\bibitem{Chang:2006eu}
  C.~H.~Chang, C.~F.~Qiao, J.~X.~Wang and X.~G.~Wu,
  Estimate of the hadronic production of the doubly charmed baryon $\Xi_{cc}$ under GM-VFN scheme,
  Phys.\ Rev.\ D {\bf 73}, 094022 (2006).

\bibitem{Chang:2007pp}
  C.~H.~Chang, J.~X.~Wang, and X.~G.~Wu,
  GENXICC: A generator for hadronic production of the double heavy baryons $\Xi_{cc}$, $\Xi_{bc}$ and $\Xi_{bb}$,
  Comput.\ Phys.\ Commun.\  {\bf 177}, 467 (2007).

\bibitem{Chang:2009va}
  C.~H.~Chang, J.~X.~Wang, and X.~G.~Wu,
  GENXICC2.0: An upgraded version of the generator for hadronic production of double heavy baryons $\Xi_{cc}$, $\Xi_{bc}$ and $\Xi_{bb}$,
  Comput.\ Phys.\ Commun.\  {\bf 181}, 1144 (2010).

\bibitem{Zhang:2011hi}
  J.~W.~Zhang, X.~G.~Wu, T.~Zhong, Y.~Yu, and Z.~Y.~Fang,
  Hadronic production of the doubly heavy baryon $\Xi_{bc}$ at LHC,
  Phys.\ Rev.\ D {\bf 83}, 034026 (2011).

\bibitem{Wang:2012vj}
  X.~Y.~Wang and X.~G.~Wu,
  GENXICC2.1: An improved version of GENXICC for hadronic production of doubly heavy baryons,
  Comput.\ Phys.\ Commun.\  {\bf 184}, 1070 (2013).

\bibitem{Chen:2014hqa}
  G.~Chen, X.~G.~Wu, J.~W.~Zhang, H.~Y.~Han, and H.~B.~Fu,
  Hadronic production of $\Xi_{cc}$ at a fixed-target experiment at the LHC,
  Phys.\ Rev.\ D {\bf 89}, 074020 (2014).


\bibitem{Li:2007vy}
  S.~Y.~Li, Z.~G.~Si, and Z.~J.~Yang,
  Doubly heavy baryon production at gamma gamma collider,
  Phys.\ Lett.\ B {\bf 648}, 284 (2007).

\bibitem{Chen:2014frw}
  G.~Chen, X.~G.~Wu, Z.~Sun, Y.~Ma, and H.~B.~Fu,
  Photoproduction of doubly heavy baryon at the ILC,
  JHEP {\bf 1412}, 018 (2014).

\bibitem{Huan-Yu:2017emk}
  H.~Y.~Bi, R.~Y.~Zhang, X.~G.~Wu, W.~G.~Ma, X.~Z.~Li, and S.~Owusu,
  Photoproduction of doubly heavy baryon at the LHeC,
  Phys.\ Rev.\ D {\bf 95}, 074020 (2017).

\bibitem{Yao:2018zze}
  X.~Yao and B.~M\"uller,
  Doubly charmed baryon production in heavy ion collisions,
  Phys.\ Rev.\ D {\bf 97}, 074003 (2018).

\bibitem{Chen:2018koh}
  G.~Chen, C.~H.~Chang, and X.~G.~Wu,
  Hadronic production of the doubly charmed baryon via the proton-nucleus and the nucleus-nucleus collisions at the RHIC and LHC,
  Eur. Phys. J. C {\bf 78}, 801 (2018).

\bibitem{Kidonakis:2004hr}
  N.~Kidonakis and R.~Vogt,
  Theoretical status of the top quark cross section,
  Int.\ J.\ Mod.\ Phys.\ A {\bf 20}, 3171 (2005).

\bibitem{Kuhn:2013zoa}
  J.~H.~K\"uhn, A.~Scharf, and P.~Uwer,
  Weak interactions in top-quark pair production at hadron colliders: An Update,
  Phys.\ Rev.\ D {\bf 91}, 014020 (2015).

\bibitem{Chang:1991bp}
  C.~H.~Chang and Y.~Q.~Chen,
  The $B(c)$ and anti-$B(c)$ mesons accessible to experiments through $Z0$ bosons decay,
  Phys.\ Lett.\ B {\bf 284}, 127 (1992).

\bibitem{Chang:1992bb}
  C.~H.~Chang and Y.~Q.~Chen,
  The production of $B(c)$ or anti-$B(c)$ meson associated with two heavy quark jets in $Z0$ boson decay,
  Phys.\ Rev.\ D {\bf 46}, 3845 (1992).

\bibitem{Braaten:1993jn}
  E.~Braaten, K.~Cheung, and T.~C.~Yuan,
  QCD fragmentation functions for $B_c$ and $B_{c}$* production,
  Phys.\ Rev.\ D {\bf 48}, R5049 (1993).

\bibitem{Braaten:1996pv}
  E.~Braaten, S.~Fleming, and T.~C.~Yuan,
  Production of heavy quarkonium in high-energy colliders,
  Annu.\ Rev.\ Nucl.\ Part.\ Sci.\  {\bf 46}, 197 (1996).

\bibitem{Lepage:1977sw}
  G.~P.~Lepage,
  A new algorithm for adaptive multidimensional integration,
  J.\ Comput.\ Phys.\  {\bf 27}, 192 (1978).

\bibitem{Bagan:1994dy}
  E.~Bagan, H.~G.~Dosch, P.~Gosdzinsky, S.~Narison, and J.~M.~Richard,
  Hadrons with charm and beauty,
  Z.\ Phys.\ C {\bf 64}, 57 (1994).

\bibitem{Field:1989uq}
  R.~D.~Field,
  Applications of perturbative QCD,
  Front.\ Phys.\  {\bf 77}, 1 (1989).

\bibitem{Altarelli:1977zs}
  G.~Altarelli and G.~Parisi,
  Asymptotic freedom in parton language,
  Nucl.\ Phys.\ B {\bf 126}, 298 (1977).


\bibitem{Falk:1993gb}
  A.~F.~Falk, M.~E.~Luke, M.~J.~Savage, and M.~B.~Wise,
  Heavy quark fragmentation to baryons containing two heavy quarks,
  Phys.\ Rev.\ D {\bf 49}, 555 (1994).

\bibitem{Braaten:1993mp}
  E.~Braaten, K.~Cheung, and T.~C.~Yuan,
  $Z^0$ decay into charmonium via charm quark fragmentation,
  Phys.\ Rev.\ D {\bf 48}, 4230 (1993).

\bibitem{Patrignani:2016xqp}
  M.~Tanabashi {\it et al.} (Particle Data Group),
  Review of Particle Physics,
  Phys.\ Rev.\ D {\bf 98}, 030001 (2018).

\bibitem{Baikov:2016tgj}
  P.~A.~Baikov, K.~G.~Chetyrkin, and J.~H.~K\"uhn,
  Five-Loop Running of the QCD Coupling Constant,
  Phys.\ Rev.\ Lett.\  {\bf 118}, 082002 (2017).

\bibitem{Herzog:2017ohr}
  F.~Herzog, B.~Ruijl, T.~Ueda, J.~A.~M.~Vermaseren, and A.~Vogt,
  The five-loop beta function of Yang-Mills theory with fermions,
  JHEP {\bf 1702}, 090 (2017).

\bibitem{Brodsky:2011ta}
  S.~J.~Brodsky and X.~G.~Wu,
  Scale setting using the extended renormalization group and the principle of maximum conformality: The QCD coupling constant at four loops,''
  Phys.\ Rev.\ D {\bf 85}, 034038 (2012).

\bibitem{Brodsky:2012rj}
  S.~J.~Brodsky and X.~G.~Wu,
  Eliminating the Renormalization Scale Ambiguity for Top-Pair Production Using the Principle of Maximum Conformality,
  Phys.\ Rev.\ Lett.\  {\bf 109}, 042002 (2012).

\bibitem{Mojaza:2012mf}
  M.~Mojaza, S.~J.~Brodsky, and X.~G.~Wu,
  Systematic All-Orders Method to Eliminate Renormalization-Scale and Scheme Ambiguities in Perturbative QCD,
  Phys.\ Rev.\ Lett.\  {\bf 110}, 192001 (2013).

\bibitem{Brodsky:2013vpa}
  S.~J.~Brodsky, M.~Mojaza, and X.~G.~Wu,
  Systematic scale-Setting to all orders: The principle of maximum conformality and commensurate scale relations,
  Phys.\ Rev.\ D {\bf 89}, 014027 (2014).

\bibitem{Wu:2013ei}
  X.~G.~Wu, S.~J.~Brodsky, and M.~Mojaza,
  The renormalization scale-setting problem in QCD,
  Prog.\ Part.\ Nucl.\ Phys.\  {\bf 72}, 44 (2013).

\bibitem{Wu:2014iba}
  X.~G.~Wu, Y.~Ma, S.~Q.~Wang, H.~B.~Fu, H.~H.~Ma, S.~J.~Brodsky, and M.~Mojaza,
  Renormalization group invariance and pptimal QCD renormalization scale-setting,
  Rep.\ Prog.\ Phys.\  {\bf 78}, 126201 (2015).

\bibitem{Sjostrand:2006za}
  T.~Sjostrand, S.~Mrenna, and P.~Z.~Skands,
  PYTHIA 6.4 physics and manual,
  JHEP {\bf 0605}, 026 (2006).

\bibitem{Bediaga:2012py}
  R.~Aaij {\it et al.} (LHCb Collaboration),
  Implications of LHCb measurements and future prospects,
  Eur.\ Phys.\ J.\ C {\bf 73}, 2373 (2013).

\bibitem{Bediaga:2018lhg}
  I.~Bediaga {\it et al.} (LHCb Collaboration),
  Physics case for an LHCb Upgrade II: Opportunities in flavour physics, and beyond, in the HL-LHC era,
  arXiv:1808.08865.

\end{thebibliography}
\end{document}